\documentstyle[tighten,preprint,aps,eqsecnum,amssymb,graphicx]{revtex}
\begin{document}
\draft
\title{Incoherent $\rho^0$ electroproduction off nuclei}
\author{T. Falter\thanks{email: Thomas.Falter@theo.physik.uni-giessen.de}, K. Gallmeister and U. Mosel}
\address{Institut fuer Theoretische Physik, Universitaet
Giessen\\ D-35392 Giessen, Germany}
\date{December 30, 2002}
\maketitle

\begin{abstract}
In the present paper we investigate incoherent $\rho^0$ electroproduction off 
complex nuclei. We derive a novel, simple expression for the incoherent 
electroproduction cross section in which one can clearly separate the final 
state interactions of the reaction products from the 'initial state 
interactions' of the photon that give rise to nuclear shadowing. In the 
special case of purely absorptive final state interactions we deduce 
from our expression the known Glauber result. A more realistic treatment of the
final state interactions within a transport model is then used to compare our
predictions with experimental data from the HERMES experiment.
\end{abstract}

\section{Introduction}\label{sec:intro}
High energy meson electroproduction off complex nuclei offers a promising tool 
to study the physics of hadron formation. The nuclear environment makes it 
possible to investigate the properties of hadrons immediately after their 
creation. In addition one can vary the energy and virtuality of the exchanged
photon to examine the phenomenon of color transparency \cite{CT}.

In the present paper we study exclusive incoherent $\rho^0$ electroproduction 
off $^{14}$N and $^{84}$Kr in the energy regime of the HERMES
experiment \cite{HERMES99,HERMES03}. In this kinematic regime the coherence 
length of the photon's $\rho^0$ component is large and quantum mechanical 
interference effects play an important role. In \cite{Fal02} we presented a 
method to account for this so-called shadowing effect in photoproduction 
within a semi-classical BUU transport model which we now apply for electron 
induced reactions. We distinguish between the 'initial 
state interactions' of the photon that give rise to shadowing and the final 
state interactions (FSI) of the produced particles. The coupled channel 
treatment of the FSI within the transport model allows for a broader spectrum 
of FSI than the usually used Glauber theory~\cite{Gla59}. In addition we 
account for Fermi motion and Pauli blocking during the reaction and take the 
finite lifetime of unstable reaction products into account.

The time that the products of an elementary photon-nucleon interaction need to
evolve to physical hadrons is called formation time. Any change in the 
reaction probability of a hadron $h$ during its formation time affects
the so-called nuclear transparency
\begin{equation}
  T_A=\frac{\sigma_{\gamma^*A\rightarrow hX}}{A\sigma_{\gamma^*N\rightarrow hX}}
\end{equation}
if the formation length, i.e. the length that $h$ travels during its formation 
time, exceeds the average internucleon distance 
($\approx 1.7$~fm) inside the nucleus.

In the case of diffractive vector meson electroproduction 
($\gamma^*N\rightarrow VN$) the virtual photon initially produces a colorless 
$q\bar{q}$-pair whose size is expected to decrease with increasing photon 
virtuality $Q^2$\cite{Kop01}. As long as the $q\bar{q}$-pair is very small, 
i.e. in the early stage of its evolution into the physical vector meson, it 
mainly reacts via its color dipole moment. This leads to a cross section that 
is quadratic in the $q\bar{q}$-pair's size. At large enough energies the 
$q\bar{q}$-pair is frozen in this small sized configuration over a distance 
that, because of time dilatation, can exceed the diameter of a nucleus. This 
effect leads to a large nuclear transparency for vector meson
production at large photon energy $\nu$ and virtuality $Q^2$ (color 
transparency). For a detailed review of color transparency see 
e.g.~\cite{CT-review} and references therein.

It is not clear whether the energy and the virtuality of the photon in the 
HERMES experiment is large enough to see an onset of color transparency. In the
present paper we do not find evidence for a finite formation time 
of the produced vector meson. The $^{14}$N data for the transparency ratio as 
function of the coherence length is compatible with the assumption that the 
diffractively produced $\rho^0$ starts interacting with a hadronic cross 
section right after its production. Kopeliovich et al.~\cite{Kop01} point 
out that this might be an accidental consequence of the specific correlation 
between the $Q^2$ of the photon and the coherence length of the $\rho^0$ in 
the HERMES data. In their work a light-cone QCD formalism was used to 
incorporate formation and coherence length effects in coherent and incoherent 
vector meson electroproduction. In addition the effect of gluon shadowing was
studied, but found to be negligible in the case of incoherent $\rho$ production.
In Ref.~\cite{Kop01} the decrease of the transparency ratio due to the
finiteness of the $\rho^0$ lifetime was discussed as well. This effect
is automatically included within our transport model. Our coupled channel 
calculation shows in addition that the effects of side-feeding in the FSI 
are unimportant because of the kinematic cuts of the HERMES experiment. 
Therefore Glauber theory seems to be appropriate for these reactions. However, 
we find that for heavy nuclei the kinematic cuts lead to an additional 
reduction of the transparency ratio due to elastic FSI which scatter the 
produced particle out of the acceptance window. One must therefore be careful 
which cross section one uses to describe the FSI in Glauber theory.

Our paper is structured in the following way: In Section~\ref{subsec:shadowing}
we show how we describe the electron-nucleon interaction and how we
account for shadowing within our model. In the limit of purely
absorptive FSI our result for the incoherent vector meson production cross
section turns out to be equivalent to the known Glauber result of 
Huefner et al.~\cite{Kop96}. However, in contrast to the result of 
Ref.~\cite{Kop96} our expression allows for an intuitive physical 
interpretation. The transport model itself is sketched in 
Section~\ref{subsec:transport}. The results of our calculations for
the exclusive incoherent $\rho^0$ production cross section within the 
coupled channel treatment are presented in Section~\ref{sec:results} in 
comparison with experimental data from the HERMES 
experiment. We close with a short summary and outlook in 
Section~\ref{sec:summary}.

\section{Model}\label{sec:model}
\subsection{Shadowing}\label{subsec:shadowing}
\begin{figure}
  \begin{center}
       \includegraphics[width=14cm]{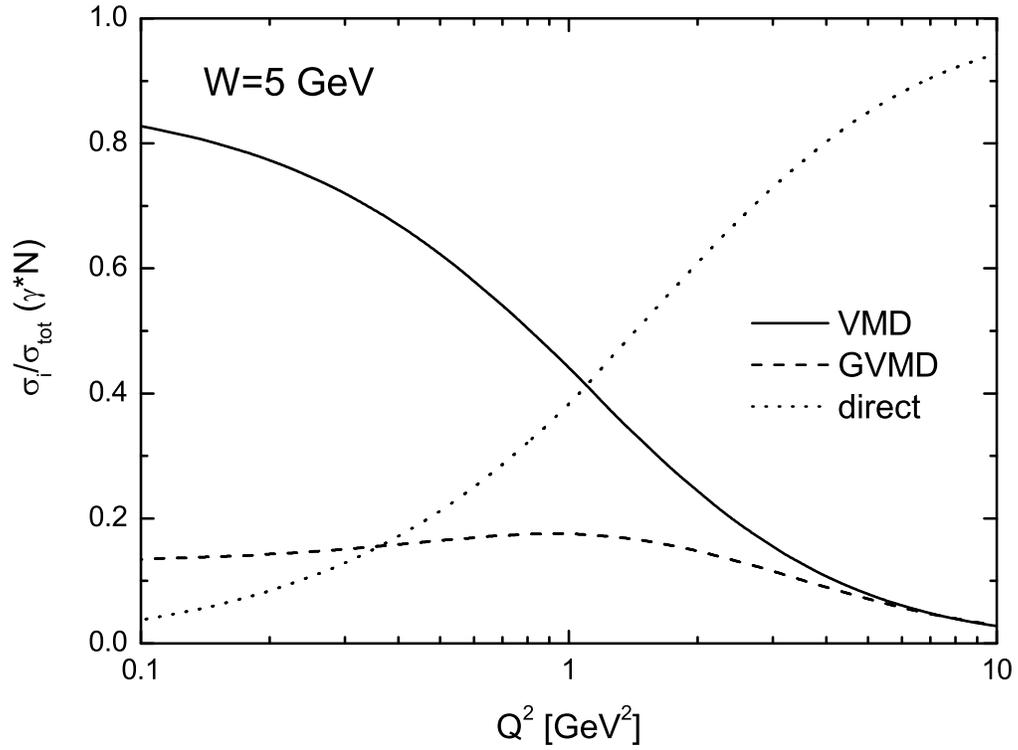}
  \end{center}
  \caption{PYTHIA result for the fraction of the total $\gamma^*N$ cross section that is contributed by the VMD (solid line), GVMD (dashed) and direct (dotted) part of the photon as a function of $Q^2$ for fixed invariant mass $W=5$~GeV. Note that in contrast to the authors of Ref.~\protect\cite{Fri00} we call every pointlike interaction of the photon a direct process.}
       \label{fig:pythia}
\end{figure}

In the one photon exchange picture the interaction of an electron and a
nucleon (or nucleus) can be reduced to the reaction of a virtual photon
on the hadronic target. The exchange of more than one photon is suppressed
by further factors of the fine structure constant $\alpha_{em}$.
We follow the method of Friberg and Sjostrand~\cite{Fri00} and use the event 
generator PYTHIA~v6.2~\cite{PYTHIA} to describe the interaction of the 
(virtual) photon and a nucleon. The basic idea is that in a photon hadron 
collision the photon not necessarily interacts as a point particle (direct 
interaction) but might fluctuate into a vector meson 
$V=\rho^0,\omega,\phi,J/\Psi$ (vector 
meson dominance) or perturbatively branch into a $q\bar{q}$ pair before the 
interaction (generalized vector meson dominance, GVMD). Within the PYTHIA model
it turns out that the latter is very unlikely 
in the kinematic regime of the HERMES experiment (photon energy 
$\nu\approx 10-20$~GeV, 
$Q^2\approx 0.5-5$~GeV$^2$) as can be seen from Fig.~\ref{fig:pythia} where
we show the contribution of the different photon components to the total
$\gamma^*N$ cross section at an invariant mass $W=$5~GeV. In most cases 
the photon will therefore
scatter deep inelastically from a parton of the target nucleon or fluctuate
into a vector meson before it reaches the nucleon. In the latter case the 
vector meson might either scatter diffractively from the nucleon or a hard 
scattering between the constituents of the vector meson and the nucleon 
might take place. The hard scattering leads to the excitation of one or two 
hadronic strings which finally fragment into hadrons. We assume that the 
hadrons that emerge from the edges of these strings, i.e. those that contain
quarks not originating from the string fragmentation, can interact with a 
hadronic cross section right after the photon nucleon reaction. Consequently,
for the diffractively produced vector meson we do not use a formation time
since none of its constituents arise from a string fragmentation.

If the struck nucleon is embedded in a nucleus one has to account for its Fermi
motion and binding energy as well as Pauli blocking of final state nucleons.
In addition one has to be aware that the nuclear environment influences
the VMD part of the photon nucleon interaction, since the vector meson 
components get modified on their way through the nuclear medium to the 
interaction point.

We express the physical photon state $|\gamma\rangle$ in terms of vector
meson states $|V\rangle$ and a state $|\gamma_0\rangle$ which consists of the
point-like photon and the GVMD part. This makes sense if one assumes that the
GVMD part does not get shadowed or as in our case is unimportant for 
kinematic reasons:
\begin{equation}
  \label{eq:vmd}
  |\gamma\rangle=\left(1-\sum_{V=\rho,\omega,\phi,J/\Psi}\frac{e^2}{2g_V^2}F_V^2\right)|\gamma_0\rangle+\sum_{V=\rho,\omega,\phi,J/\Psi}\frac{e}{g_V}F_V|V\rangle.
\end{equation}
The formfactor 
\begin{equation}
  \label{eq:formfactor}
  F_V^2=\left(\frac{W^2}{Q^2+W^2}\right)^3\left[1+0.5\frac{4m_V^2Q^2}
{(m_V^2+Q^2)^2}\right]\left(\frac{m_V^2}{m_V^2+Q^2}\right)^2
\end{equation}
is taken from~\cite{Fri00} and also accounts for contributions of longitudinal 
photons. In Eq. (\ref{eq:formfactor}) $\nu$ denotes the energy of the 
photon, $Q^2$ its virtuality and $m_V$ the mass of the vector meson $V$. $W$ is
the invariant mass of the photon nucleon system.

For a photon with momentum $k\vec{e}_z$ the wave function of the component $V$ 
can be written as
\begin{equation}
  \label{eq:psiV0}
  \psi_V^{(0)}(\vec b,z)=\chi(\vec b)e^{ikz}.
\end{equation}
On the way through the nucleus the vector meson components get modified. 
Consider an ensemble of $A$ nucleons at positions $\{ \vec r_i\}=\{(\vec s_i,z_i)\}$ 
which are labeled in such a way that $z_1<z_2<...<z_A$. According 
to Glauber theory~\cite{Yen71} the wave function $\psi_V$ behind the first 
nucleon looks like
\begin{eqnarray}
  \label{eq:psiV1}
  \psi_V^{(1)}(\vec b,z)&=&\psi_V^{(0)}(\vec b,z)-\Gamma_V(\vec b-\vec s_1)\psi_V^{(0)}(\vec b,z_1)e^{ik_V(z-z_1)}\\ \nonumber
  &=&\chi(\vec b)\left(e^{ikz}-\Gamma_V(\vec b-\vec s_1)e^{iq_Vz_1}e^{ik_Vz}\right)
\end{eqnarray}
if one neglects off-diagonal scattering from one vector meson component into
another (diagonal approximation). The phase factor in the second term arises 
from putting the vector meson component on its mass shell, i.e. 
$k_V=\sqrt{\nu^2-m_V^2}$ and $q_V=k-k_V$ is the corresponding momentum 
transfer. The profile function $\Gamma_V$ is related to the elastic vector 
meson nucleon scattering amplitude $f_V$ in the following way:
\begin{equation}
  \label{eq:profile}
  \Gamma_V(\vec b)=\frac{1}{2\pi i\vec k_V}\int e^{i\vec k_t\cdot\vec b}f_V(\vec k_t)d^2k_t.
\end{equation}
After passing the second nucleon the wave function takes on the form
\begin{equation}
  \label{eq:psiV2}
  \psi_V^{(2)}(\vec b,z)=\chi(\vec b)\left(e^{ikz}-\Gamma_V(\vec b-\vec s_2)e^{iq_Vz_2}e^{ik_Vz}-\Gamma_V(\vec b-\vec s_1)(1-\Gamma_V(\vec b-\vec s_2)e^{iq_Vz_1}e^{ik_Vz}\right).
\end{equation}
At position $z$ the whole ensemble of $A$ nucleons has led to the modification
\begin{eqnarray}
  \psi_V^{(A)}(\vec b,z)&=&\chi(\vec b)\left(e^{ikz}-\sum_{j=1}^A\Gamma_V(\vec b-\vec s_j)e^{iq_Vz_j}\prod_{k\neq j}^A\left[1-\Gamma_V(\vec b-\vec s_k)\Theta(z_k-z_j)\right]\Theta(z-z_k)e^{ik_Vz}\right) \nonumber \\
    &=&\left\{1-\sum_{j=1}^A\Gamma_V(\vec b-\vec s_j)e^{iq_V(z_i-z)}\prod_{k\neq j}^A\left[1-\Gamma_V(\vec b-\vec s_k)\Theta(z_k-z_j)\right]\Theta(z-z_k)\right\}\psi_V^{(0)}(\vec b,z) \nonumber \\
    &=&\left\{1-\Gamma^{(A)}_V(\vec b,z;\{\vec r_i\})\right\}\psi_V^{(0)}(\vec b,z)  \label{eq:psiVn}.
\end{eqnarray}
Hence, the physical photon state changes on its way through the nucleus to 
position $\vec r=(\vec b,z)$ according to
\begin{equation}
\label{eq:inmediumstate}
  |\gamma(\vec r)\rangle=\left(1-\sum_{V=\rho,\omega,\phi,J/\Psi}\frac{e^2}{2g_V^2}F_V^2\right)|\gamma_0\rangle+\sum_{V=\rho,\omega,\phi,J/\Psi}\frac{e}{g_V}F_V\left(1-\overline{\Gamma^{(A)}_V}(\vec r)\right)|V\rangle
\end{equation}
where the nuclear profile function from (\ref{eq:psiVn}) must be averaged 
over the positions of the nucleons in the nucleus
\begin{eqnarray}
  \overline{\Gamma^{(A)}_V}(\vec b,z)&=&\langle0|\Gamma^{(A)}_V(\vec b,z;\{\vec r_i\})|0\rangle \nonumber\\
  &=&\intop_{-\infty}^{z}dz_in(\vec b,z_i)\frac{\sigma_{VN}}{2}(1-i\alpha_V)e^{iq_V(z_i-z)}\exp\left[-\frac{1}{2}\sigma_{VN}(1-i\alpha_V)\intop_{z_i}^{z}dz_kn(\vec b,z_k)\right].  \label{eq:av_profile}
\end{eqnarray}
To get to the second line of Eq. (\ref{eq:av_profile}) we have followed the way
of \cite{Yen71}, i.e. making use of the independent particle model and the 
large $A$ limit and used the optical theorem to express the elastic forward 
scattering amplitude in terms of the total $VN$ cross section $\sigma_V$:
\begin{equation}
  \label{eq:optical}
  f_V(\vec 0)=\frac{ik_V}{4\pi}\sigma_V(1-i\alpha_V).
\end{equation}
The nucleon number density is denoted by $n(\vec r)$ and normalized to $A$.

Neglecting any influence of the FSI, the reaction amplitude for the process
$\gamma N\rightarrow f$ on a nucleon at position $\vec r$ inside a nucleus 
changes compared to the vacuum due to these 'initial state interactions' of the
photon: 
\begin{equation}
  \label{eq:transition}
  \langle f|\hat{T}|\gamma\rangle\rightarrow\langle f|\hat{T}|\gamma(\vec r)\rangle.
\end{equation}
By comparing Eq. (\ref{eq:vmd}) and (\ref{eq:inmediumstate}) one sees that to
account for shadowing in incoherent photoproduction one only has to 
multiply the amplitude of each vector meson states $|V\rangle$ by the 
corresponding factor $\left(1-\overline{\Gamma^{(A)}_V}(\vec r)\right)$. This 
is equivalent to the method we used in~\cite{Fal02}.

We now compare our method of describing shadowing of incoherent 
electroproduction with conventional Glauber results. We therefore calculate 
the incoherent vector meson electroproduction cross section off a nucleus and 
assume purely absorptive FSI for the moment. After averaging over all 
positions $\vec r=(\vec b,z)$ we get:
\begin{eqnarray}
  \label{eq:falter}
  \sigma_{\gamma A\rightarrow VA^*}&=&\sigma_{\gamma N\rightarrow VN}\intop d^2b\intop_{-\infty}^{\infty}dz n(\vec b,z)\nonumber\\
  & &\times\left|1-\intop_{-\infty}^{z}dz_i n(\vec b,z_i)\frac{\sigma_V}{2}(1-i\alpha_V)e^{iq_V(z_i-z)}\exp\left[-\frac{1}{2}\sigma_V(1-i\alpha_V)\intop_{z_i}^{z}dz_kn(\vec b,z_k)\right]\right|^2\nonumber\\
  & &\quad\times \exp\left[-\sigma_V^{inel}\intop_{z}^\infty dz' n(\vec b,z')\right].
\end{eqnarray}
In (\ref{eq:falter}) we again neglect off-diagonal scattering, i.e. $V$ 
production is triggered by the $V$ component of the photon only. The last 
factor in (\ref{eq:falter}) includes the inelastic $VN$ cross section 
$\sigma_V^{inel}$ and accounts for the FSI. 

The production process described by Eq.(\ref{eq:falter}) can be 
understood as follows: The incoherent vector meson production takes place on a 
nucleon at position $(\vec b,z)$. The production is triggered either directly 
by the photon or via an intermediate vector meson which was earlier produced on
a nucleon at position $(\vec b,z_i)$ without excitation of the nucleus. The 
interactions of this vector meson on its way from $z_i$ to position $z$ are of
optical potential type and leave the nucleus in its ground state. This 
propagation is described by the second exponential in (\ref{eq:falter}). The 
interference between the direct and the indirect process leads to shadowing.
The nucleus gets excited in the incoherent reaction at position $z$. The
possibility that the vector meson is lost on its way from position $z$ 
out of the nucleus is taken care of by the last exponential.

Expression (\ref{eq:falter}) formally differs only slightly from the approximate result 
for the incoherent vector meson production cross section which is given without
derivation in Ref.~\cite{Yen71}. The latter can be obtained by making the
following replacements in (\ref{eq:falter}):
\begin{equation}
  \label{eq:replacements}
  e^{iq_V(z_i-z)}\rightarrow e^{iq_Vz_i},\qquad\sigma_V^{inel}\rightarrow\sigma_V.
\end{equation}
The last replacement simply means that one neglects the possibility of elastic 
$VN$ scattering whereas the occurrence of the different phase factor is 
unclear. The Authors of Ref.~\cite{Kop96} have given a seemingly different 
result for incoherent $V$ electroproduction:
\begin{eqnarray}
  \label{eq:kopeliovich}
  \sigma_{\gamma A\rightarrow VA^*}&=&\sigma_{\gamma N\rightarrow VN}\int d^2b\biggl\{\intop_{-\infty}^{+\infty}dzn(\vec b,z)e^{-\sigma_{V}^{inel}T_z(\vec b)}\nonumber\\
  &  &\quad+\frac{1}{2}\frac{\sigma_{V}}{\sigma_{V}^{el}}(\sigma_{V}^{inel}-\sigma_{V}^{el})\intop_{-\infty}^{+\infty}dz_1n(\vec b,z_1)\intop_{z_1}^{+\infty}dz_2n(\vec b,z_2)\nonumber\\
  &  &\qquad\times\cos\left[q_V(z_1-z_2)\right]e^{-\frac{1}{2}(\sigma_{V}^{inel}-\sigma_{V}^{el})T_{z_2}(\vec b)-\frac{1}{2}\sigma_{V}T_{z_1}(\vec b)}\nonumber\\
  & &\quad-\frac{1}{4}\frac{(\sigma_{V})^2}{\sigma_{V}^{el}}\left|\intop_{-\infty}^{+\infty}dzn(\vec b,z)e^{iq_Vz}e^{-\frac{1}{2}\sigma_{V}T_z(\vec b)}\right|^2\biggl\}
\end{eqnarray}
with $T_z(\vec b)=\intop_z^{\infty}n(\vec b,z')dz'$ and 
$T(\vec b)=T_{z=-\infty}(\vec b)$. Expression (\ref{eq:kopeliovich}) can be 
interpreted as follows: The last term represents the coherent $V$
photoproduction cross section which is subtracted from the inclusive $V$ 
photoproduction cross section to yield the incoherent part. In Appendix A we 
show that the physically more transparent expression (\ref{eq:falter}) is 
mathematically identical to (\ref{eq:kopeliovich}).

\subsection{Transport model}\label{subsec:transport}
Up to now we have ignored the effects of Fermi motion, binding energies and 
Pauli blocking of final state nucleons in the process described by 
(\ref{eq:transition}). In addition we neglected the finite life time of the 
$\rho^0$ and took only absorptive FSI into account. These shortcomings are 
avoided when using the model of Ref.~\cite{Fal02,Eff99}. The propagation of the 
final state $|f\rangle$ through the nucleus is treated within a semi-classical
transport model based on the Boltzmann-Uehling-Uhlenbeck (BUU) equation.
The BUU equation describes the time evolution of the phase space density 
$f_i(\vec r,\vec p,t)$ of particles of type $i$ that can interact via binary 
reactions. Besides the nucleons these particles involve baryonic 
resonances and mesons ($\pi$, $\eta$, $\rho$, $K$, ...) that are produced 
either in the primary reaction or during the FSI. For a particle species $i$
the BUU equation can be written as:
\begin{equation}
  \left(\frac{\partial}{\partial t}+\frac{\partial H}{\partial\vec r}\frac{\partial}{\partial \vec r}-\frac{\partial H}{\partial \vec r}\frac{\partial}{\partial \vec p}\right)f_i(\vec r,\vec p,t)=I_{coll}[f_1,...f_i,...,f_M].
\end{equation}
For baryons the Hamilton function $H$ includes a mean field potential which 
depends on the particle position and momentum. The collision integral on the 
right hand side accounts for the creation and
annihilation of particles of type $i$ in a collision as well as elastic 
scattering from one position in phase space into another. For fermions Pauli 
blocking is taken into account in $I_{coll}$ via blocking factors. The BUU 
equations of each particle species $i$ are coupled via the mean 
field and the collision integral. The resulting system of coupled 
differential-integral equations is solved via a test particle ansatz for 
the phase space density. For details of the transport model see 
Ref.~\cite{Eff99}.

The classes of FSI that are included in the transport model goes
far beyond what can be achieved within Glauber theory. As a result the finally
observed $\rho^0$ does not need to be created in the primary reaction but might
be produced during the FSI via side feeding. It is therefore clear that a purely 
absorptive treatment of the FSI as in Glauber theory can only be used if
one is sure that one has eliminated the possibility of side feeding by 
applying enough constraints on the observable (see~\cite{Fal02} for details).

We also stress that within our model instable particles might decay 
during their propagation through the nucleus. In case of the $\rho^0$ this 
means that both pions have to escape the nucleus without further rescattering 
to make the identification of the $\rho^0$ still possible.

\section{Results}\label{sec:results}
\begin{figure}
  \begin{center}
    \includegraphics[width=14cm]{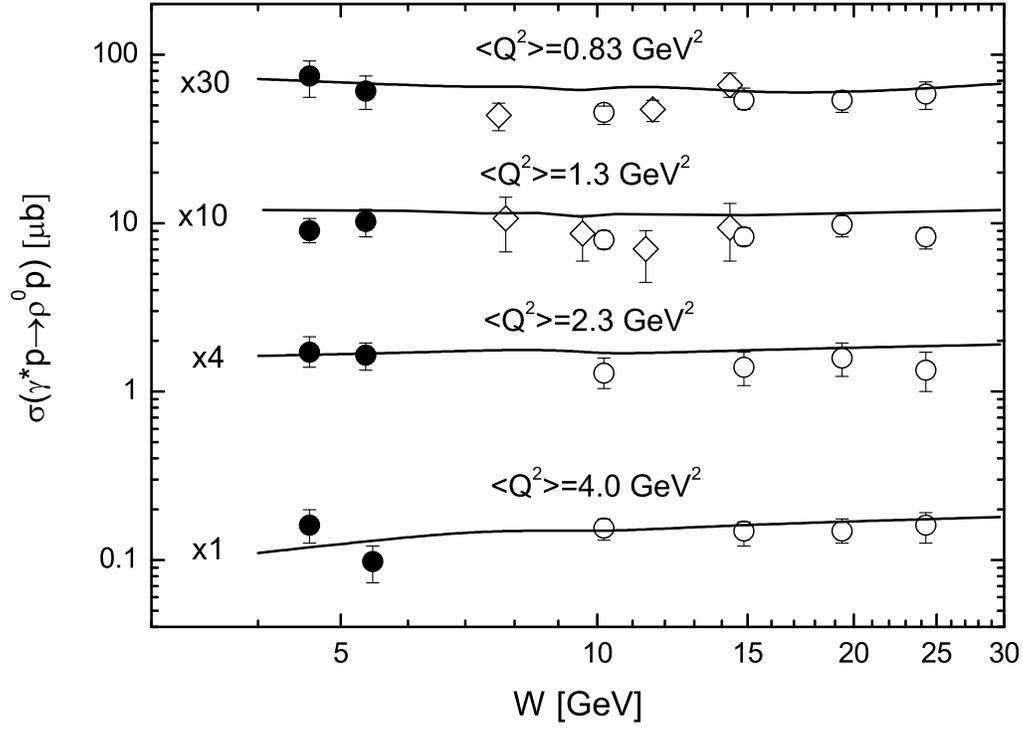}
  \end{center}
  \caption{The virtual-photoproduction cross section for $\rho^0$ production plotted versus the invariant mass $W$ at average $Q^2$ values of $0.83$, $1.3$, $2.3$ and $4.0$ GeV$^2$. The data has been taken from the HERMES \protect\cite{HERMES00} (filled circles), E665 \protect\cite{E665} (open circles) and CHIO collaboration \protect\cite{CHIO} (open diamonds).}
  \label{fig:elementary}
\end{figure}

\begin{figure}
  \begin{center}
    \includegraphics[width=14cm]{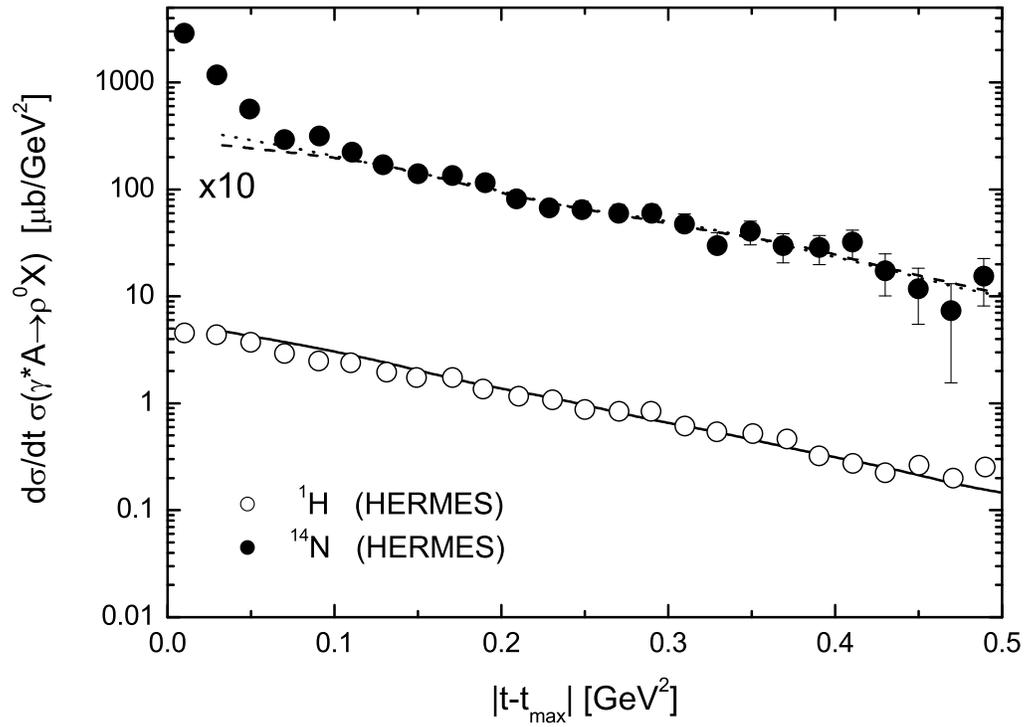}
  \end{center}
  \caption{The exclusive incoherent differential $\rho^0$ production cross section for $^1H$ (solid line) and $^{14}$N (dashed line) in comparison with experimental data from the HERMES collaboration~\protect\cite{HERMES99}. The calculation has been performed for $\nu=13$~GeV and $Q^2=1.7$~GeV$^2$. The dotted curve represents a calculation for $^{14}$N without Pauli blocking and Fermi motion. For details on the exclusivity measure see text.} 
  \label{fig:dsigdt}
\end{figure}

Before we turn to nuclear targets we first verify that the input of our
model is reasonable. We therefore look at exclusive $\rho^0$ production
off hydrogen. We use the same kinematical cuts as in the HERMES experiment~
\cite{HERMES00}, i.e. the final state has to consist of two oppositely charged 
pions with invariant mass between $0.6$~GeV and $1$~GeV. The four momentum 
transfer $|t-t_{max}|$ between the virtual photon and the $\pi^+\pi^-$-pair 
has to be smaller than 0.4~GeV$^2$ and we apply the exclusivity measure 
\begin{equation}
  \label{eq:exclms}
  \Delta E=\frac{p_Y^2-m_N^2}{2m_N}<0.4\text{~GeV},
\end{equation}
where $m_N$ denotes the nucleon mass and 
\begin{equation}
  \label{eq:py}
  p_Y=p_N+p_{\gamma}-p_{\rho}
\end{equation}
the 4-momentum of the undetected final state. In Eq. (\ref{eq:py}) 
$p_{\gamma}$ and $p_{\rho}$ denote the 4-momenta of the incoming photon and 
the detected $\pi^+\pi^-$ pair and $p_N$ is the 4-momentum 
of the struck nucleon which, for the calculation of $p_Y$, is assumed to be at 
rest. In Fig.~\ref{fig:elementary} we compare our calculation of the exclusive 
$\rho^0$ production cross section off hydrogen with experimental 
data~\cite{HERMES00,E665,CHIO}. In the whole $Q^2$ region covered by the
HERMES experiment we find very good agreement for a broad range of the 
invariant mass $W$ of the photon nucleon system. Also the slope of the 
differential production cross section is reproduced very well by the PYTHIA 
model as can be seen from the solid line in Fig.~\ref{fig:dsigdt} where we 
show our calculation of $\frac{d\sigma}{dt}(\gamma^*p\rightarrow\rho^0p)$ 
together with the HERMES data \cite{HERMES99}.

For our calculations on exclusive $\rho^0$ electroproduction off nuclei we 
again use the kinematic cuts of the HERMES collaboration 
\cite{HERMES99,HERMES03}. This 
means that we restrict our exclusivity measure to the region
\begin{equation}
  -2\text{~GeV}<\Delta E<0.6\text{~GeV},
\end{equation}
and introduce a lower boundary for the four-momentum transfer 
$|t-t_{max}|>$0.09~GeV$^2$ as imposed by the HERMES collaboration to get rid
of coherently produced $\rho^0$. From the dashed line in Fig.~\ref{fig:dsigdt} 
one sees that the differential $\rho^0$ electroproduction cross section off 
$^{14}$N is again in excellent agreement with the HERMES data \cite{HERMES99}. 
Throughout our calculations the effect of the nucleon potential turns out to 
be negligible. This is reasonable since the involved energies
are much larger than the typical binding energies which are in the order of a 
few MeV. A combined effect of Fermi motion and Pauli blocking on the 
incoherent differential production cross section is visible at 
$|t-t_{max}|<0.1$~GeV$^2$. At $|t-t_{max}|\approx0.05$~GeV$^2$ the differential
cross section is about 25\% smaller than the calculation
without Fermi motion and Pauli blocking (dotted curve). The reason for
this reduction is that in the case that the bound nucleon moves towards the 
incoming photon the outgoing nucleon might be Pauli blocked for small momentum 
transfers. Note that this Pauli blocking just means that the whole nucleus 
absorbs the transferred momentum, i.e. this event contributes to the coherent 
production cross section which we do not consider here.

\begin{figure}
  \begin{center}
    \includegraphics[width=14cm]{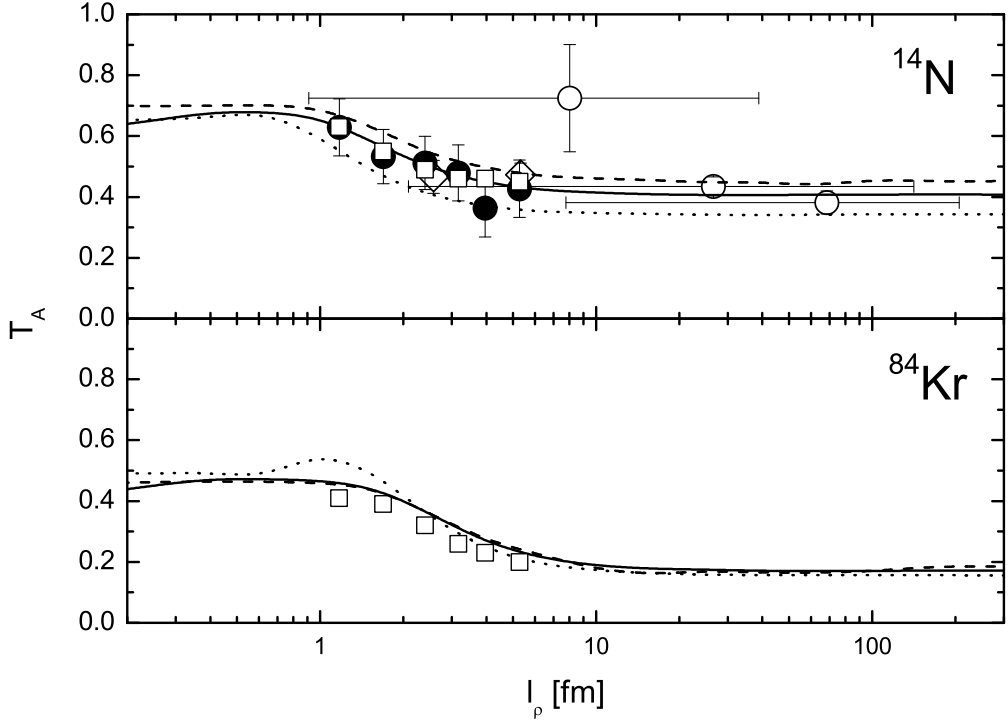}
  \end{center}
  \caption{Nuclear transparency ratio $T_A$ for $\rho^0$ electroproduction
plotted versus the coherence length of the $\rho^0$ component of the photon.
The data is taken from~\protect\cite{HERMES03} (filled circles), 
\protect\cite{Ada95} (open circles) and \protect\cite{Cle69} (open diamonds).
The solid line represents the Glauber result when using (\ref{eq:falter}) 
including nucleon-nucleon correlations, the dotted line shows the effect of the
substitutions (\ref{eq:replacements}). The dashed curve represents the Glauber
result using (\ref{eq:kopeliovich}) with the density distribution of 
Ref.~\protect\cite{Kop96}. For each transparency ratio calculated within our 
transport model (open squares) we used the average value of $Q^2$ and $\nu$ of the corresponding HERMES data point.}
\label{fig:rho}
\end{figure}

In Fig.~\ref{fig:rho} we show the transparency ratio $T_A$ for exclusive 
$\rho^0$ production as a function of the coherence length $l_\rho=q_\rho^{-1}$.
The solid line is the result that one gets if one uses Eq. (\ref{eq:falter})
and accounts for two-body correlations by making the 
substitution~\cite{Fal00}
\begin{equation}
  n(\vec b,z_i)\rightarrow n(\vec b,z_i)(1-j_0(q_c|z_i-z|))
\end{equation}
in (\ref{eq:falter}) with $q_c=0.78$~GeV. This is necessary to avoid 
unphysical contributions from processes where $z_i\approx z$ which would 
contribute for small values of the coherence length $l_\rho$. In 
Ref.~\cite{Fal00} we showed that this Bessel function parameterization yields a
good description of shadowing in photoabsorption. In Eq. (\ref{eq:falter}) we 
use for the total $\rho^0N$ cross section $\sigma_{\rho^0}=25$~mb and for the 
elastic part $\sigma_{el}=3$ mb. These two values correspond to the $\rho N$ 
cross sections used  within the transport model for the involved $\rho^0$ 
momenta. 

The result of the transport model is represented by the open 
squares. For each HERMES data point \cite{HERMES03} we have made a 
separate calculation with the corresponding $\nu$ and $Q^2$. In the case of 
$^{14}$N the Glauber and the 
transport calculation are in perfect agreement with each other and the 
experimental data. This demonstrates that, as we have discussed in 
Ref.~\cite{Fal02}, Glauber theory can be used for the FSI if the right 
kinematic constraints are applied. For comparison we also 
show the result of Huefner et al.~\cite{Kop96} (dashed curve) as well as the 
result that one gets when using the approximate expression by 
Yennie~\cite{Yen71} including two-body correlations (dotted curve). The 
somewhat larger transparency ratio of Huefner et al. arises from the different 
density distribution that the authors of Ref.~\cite{Kop96} use which reduces 
the effect of FSI. The difference of
the Yennie result (dotted curve) and our calculation arises mainly from the 
different cross section in the FSI (second replacement in 
(\ref{eq:replacements})). The different phase factor leads to the change in the
transparency ratio at small coherence lengths, since $q_V=l_V^{-1}$.

After applying all of the above cuts, nearly all of the detected $\rho^0$ 
stem from diffractive $\rho^0$ production for which the formation time is zero.
The $^{14}$N data seems to support the assumption that the time needed 
to put the preformed $\rho^0$ fluctuation on its mass shell and let the wave 
function evolve to that of a physical $\rho^0$ is small for the considered 
values of $Q^2$. Furthermore, the photon energy is too low to yield a large 
enough $\gamma$ factor to make the formation length exceed the internucleon 
distance and make color transparency visible. This conclusion is at variance
with that reached in Ref.~\cite{Kop01}.

We now turn to $^{84}$Kr where we expect a stronger effect of the FSI. 
Unfortunately there is yet no data available to compare with. As can be seen
from Fig.~\ref{fig:rho} the transport calculation for $^{84}$Kr gives a 
slightly smaller transparency ratio than the Glauber calculation, especially at
low values of the coherence length, i.e. small momenta of the produced 
$\rho^0$. There are two reasons for this: About 10\%~of the difference arises 
from the fact that within the transport model the $\rho^0$ is allowed to decay
into two pions. The probability that at least one of the pions interacts on 
its way out of the nucleus is about twice as large as that of the $\rho^0$. 
The other reason is that in the Glauber calculation (\ref{eq:falter}) only 
the inelastic part of the $\rho^0N$ cross section enters whereas the transport
calculation contains the elastic part as well. Thus all elastic scattering
events out of the experimentally imposed $t$-window are neglected in the
Glauber description. It is because of this $t$-window that also elastic 
$\rho^0N$ scattering reduces the transport transparency ratio shown in 
Fig.~\ref{fig:rho}. Both effects are more enhanced at lower energies and become
negligible for the much smaller $^{14}$N nucleus.

\section{Summary and Outlook}\label{sec:summary}
We have developed a method to account for coherence length effects in 
incoherent electroproduction off nuclei which allows us to distinguish between 
the initial state interactions of the photon (shadowing) and the FSI of the 
reaction products. We have shown that our result is equivalent to the exact 
result of Glauber theory if one treats the FSI as a purely absorptive one. We 
have then 
performed a coupled channel treatment of the FSI within a semi-classical 
transport model and calculated the transparency ratio for exclusive incoherent 
$\rho^0$ photoproduction off $^{14}$N and $^{84}$Kr. The result for $^{14}$N is
in agreement with experimental data and with the Glauber prediction. The 
latter shows that in the case of $^{14}$N Glauber theory is applicable after 
the kinematic cuts of the HERMES experiment are applied. Since we do not use a 
formation time for diffractively produced vector mesons we deduce that one 
cannot see an onset of color transparency in the nitrogen data. For the 
$^{84}$Kr target no experimental data is available to compare with. 
However, we find deviations from the simple Glauber model because of the 
finite life time of the $\rho^0$ and elastic scattering out of the 
kinematically allowed $|t|$-region. These effects should be taken into account 
when evaluating the $^{84}$Kr data in search of color transparency.
As discussed by Kopeliovich et al.~\cite{Kop01} one might see an onset of color
transparency when investigating the transparency ratio as a function of $Q^2$ 
for fixed coherence length \cite{HERMES03}. Including a formation time for 
exclusive $V$ production is not straightforward within our model since a change
of the $VN$ cross section during the vector meson formation will influence both
the initial as well as the final state interactions. This is planned for future
work, when hopefully also the $^{84}$Kr data has become available.

\section*{Acknowledgments}
The authors want to thank A. Borissov and W. Cassing for useful discussions. 
This work was supported by DFG.

\appendix
\section*{A}
In the following we show the equality of Eqs. (\ref{eq:falter}) and 
(\ref{eq:kopeliovich}). Since the real part of the $VN$ scattering amplitude 
was neglected in the derivation of (\ref{eq:kopeliovich}) we also set 
$\alpha_V=0$ in Equation (\ref{eq:falter}) and end up with: 
\begin{eqnarray}
\label{eq:appendix1}
  \sigma_{\gamma A\rightarrow VA^*}&=&\sigma_{\gamma N\rightarrow VN}\intop d^2b\intop_{-\infty}^{\infty}dz n(\vec b,z)e^{-\sigma_V^{inel}\intop_{z}^\infty dz' n(\vec b,z')}\nonumber\\
  & &\times\left|1-\intop_{-\infty}^{z}dz_i n(\vec b,z_i)\frac{\sigma_V}{2}e^{iq_V(z_i-z)}\exp\left[-\frac{1}{2}\sigma_V\intop_{z_i}^{z}dz_kn(\vec b,z_k)\right]\right|^2\nonumber\\
  &=&\sigma_{\gamma N\rightarrow VN}\intop d^2b\intop_{-\infty}^{\infty}dz n(\vec b,z)e^{-\sigma_V^{inel}\intop_{z}^\infty dz' n(\vec b,z')}\nonumber\\
  & &\times\biggl\{1-\sigma_V\intop_{-\infty}^zdz_in(\vec b,z_i)\cos{\left[q_V(z_i-z)\right]\exp\left[-\frac{1}{2}\sigma_V\intop_{z_i}^{z}dz_kn(\vec b,z_k)\right]}\nonumber\\
  & &+\frac{\sigma_V^2}{4}\left|\intop_{-\infty}^{z}dz_i n(\vec b,z_i)e^{iq_Vz_i}\exp\left[-\frac{1}{2}\sigma_V\intop_{z_i}^{z}dz_kn(\vec b,z_k)\right]\right|^2\biggl\}\nonumber\\
  &=&\sigma_{\gamma N\rightarrow VN}\intop d^2b\biggl\{\intop_{-\infty}^{\infty}dz n(\vec b,z)e^{-\sigma_V^{inel}T_z(\vec b)}\nonumber\\
  & &-\sigma_V\intop_{-\infty}^{\infty}dz_1n(\vec b,z_1)\intop_{z_1}^{\infty}dz_2n(\vec b,z_2)\cos\left[q_V(z_1-z_2)\right]\nonumber\\
  & &\times\exp\left[-\sigma_V^{inel}\intop_{z_2}^\infty dz' n(\vec b,z')-\frac{1}{2}\sigma_V\intop_{z_1}^{z_2}dz'n(\vec b,z')\right]\nonumber\\
  & &+\frac{\sigma_V^2}{4}\intop_{-\infty}^{\infty}dzn(\vec b,z)\nonumber\\
  & &\times\left|\intop_{-\infty}^{z}dz_in(\vec b,z_i)e^{iq_Vz_i}\exp\left[-\frac{1}{2}\sigma_V\intop_{z_i}^zdz'n(\vec b,z')-\frac{1}{2}\sigma_V^{inel}\intop_z^{\infty}dz'n(\vec b,z')\right]\right|^2\biggl\}.
\end{eqnarray}
In the second step we have renamed the integration variables and rewritten the 
integral limits. The first term of (\ref{eq:appendix1}) already equals that the
first term in (\ref{eq:kopeliovich}). The exponent in the second term of 
(\ref{eq:appendix1}) yields
\begin{equation}
  -\sigma_V^{inel}T_{z_2}(\vec b)-\frac{1}{2}\sigma_V\left(T_{z_1}(\vec b)-T_{z_2}(\vec b)\right)=-\frac{1}{2}\left(\sigma_V^{inel}-\sigma_V^{el}\right)T_{z_2}(\vec b)-\frac{1}{2}\sigma_VT_{z_1}(\vec b)
\end{equation}
and the exponent of the third term in (\ref{eq:appendix1}) is
\begin{equation}
  -\frac{1}{2}\left(\sigma_VT_{z_i}(\vec b)-T_{z}(\vec b)\right)-\frac{1}{2}\sigma_V^{inel}T_{z}(\vec b)=-\frac{1}{2}\sigma_VT_{z_i}(\vec b)+\frac{1}{2}\sigma_V^{el}T_{z}(\vec b).
\end{equation}
We now further manipulate the last term of (\ref{eq:appendix1}):
\begin{eqnarray}
\label{eq:appendix2}
  \frac{\sigma_V^2}{4}& &\intop_{-\infty}^{\infty}dzn(\vec b,z)e^{\sigma_V^{el}T_z(\vec b)}\left|\intop_{-\infty}^{z}dz_in(\vec b,z_i)e^{iq_Vz_i}e^{-\frac{1}{2}\sigma_VT_{z_i}(\vec b)}\right|^2\nonumber\\
  & &=\frac{\sigma_V^2}{4}\intop_{-\infty}^{\infty}dz\intop_{-\infty}^{z}dz_1\intop_{-\infty}^{z}dz_2n(\vec b,z)n(\vec b,z_1)n(\vec b,z_2)e^{\sigma_V^{el}T_z(\vec b)}\cos\left[q_V(z_1-z_2)\right]e^{-\frac{1}{2}\sigma_V(T_{z_1}(\vec b)+T_{z_2}(\vec b))}.
\end{eqnarray}
Rewriting the integral in the following form
\begin{equation}
  \intop_{-\infty}^{\infty}dz\intop_{-\infty}^{z}dz_1\intop_{-\infty}^{z}dz_2=\intop_{-\infty}^{\infty}dz_2\intop_{-\infty}^{z_2}dz_1\intop_{z_2}^{\infty}dz+\intop_{-\infty}^{\infty}dz_1\intop_{-\infty}^{z_1}dz_2\intop_{z_1}^{\infty}dz
\end{equation}
and using the symmetry of the integrand with respect to the variables $z_1$ and $z_2$ we can rewrite (\ref{eq:appendix2}) as
\begin{eqnarray}
  \label{eq:appendix3}
  \frac{\sigma_V^2}{2}& &\intop_{-\infty}^{\infty}dz_2n(\vec b,z_2)\intop_{-\infty}^{z_2}dz_1n(\vec b,z_1)\cos\left[q_V(z_1-z_2)\right]e^{-\frac{1}{2}\sigma_V(T_{z_1}(\vec b)+T_{z_2}(\vec b))}\intop_{z_2}^{\infty}dzn(\vec b,z)e^{\sigma_V^{el}T_z(\vec b)}\nonumber\\
  & &=-\frac{1}{2}\frac{\sigma_V^2}{\sigma_V^{el}}\intop_{-\infty}^{\infty}dz_2n(\vec b,z_1)\intop_{-\infty}^{z_2}dz_1n(\vec b,z_2)\cos\left[q_V(z_1-z_2)\right]e^{-\frac{1}{2}\sigma_V(T_{z_1}(\vec b)+T_{z_2}(\vec b))}\left(1-e^{\sigma_V^{el}T_{z_2}}\right)\nonumber\\
  & &=-\frac{1}{2}\frac{\sigma_V^2}{\sigma_V^{el}}\intop_{-\infty}^{\infty}dz_1n(\vec b,z_1)\intop_{z_1}^{\infty}dz_2n(\vec b,z_2)\cos\left[q_V(z_1-z_2)\right]e^{-\frac{1}{2}\sigma_V(T_{z_1}(\vec b)+T_{z_2}(\vec b))}\nonumber\\
  & &\quad+\frac{1}{2}\frac{\sigma_V^2}{\sigma_V^{el}}\intop_{-\infty}^{\infty}dz_1n(\vec b,z_1)\intop_{z_1}^{\infty}dz_2n(\vec b,z_2)\cos\left[q_V(z_1-z_2)\right]e^{-\frac{1}{2}(\sigma_V^{inel}-\sigma_V^{el})T_{z_2}(\vec b)-\frac{1}{2}\sigma_VT_{z_1}(\vec b))}.
\end{eqnarray}
In the first step we have performed the integral over $z$. In the last 
step we have again rewritten the integral over $z_1$ and $z_2$ to be able to 
combine the second term of (\ref{eq:appendix3}) with the second term of 
(\ref{eq:appendix1}) to get the second term in (\ref{eq:kopeliovich}). Now we 
only have to show the equality of the last
term in (\ref{eq:kopeliovich}) and the first one in (\ref{eq:appendix3}). 
Therefore we rewrite 
\begin{eqnarray}
  -\frac{1}{4}& &\frac{\sigma_V^2}{\sigma_V^{el}}\left|\intop_{-\infty}^{\infty}dzn(\vec b,z)e^{iq_Vz}e^{-\frac{1}{2}\sigma_VT_z(\vec b)}\right|^2\nonumber\\
  & &=-\frac{1}{4}\frac{\sigma_V^2}{\sigma_V^{el}}\intop_{-\infty}^{\infty}dz_1\intop_{-\infty}^{\infty}dz_2n(\vec b,z_1)n(\vec b,z_2)\cos\left[q_V(z_1-z_2)\right]e^{-\frac{1}{2}\sigma_V(T_{z_1}(\vec b)+T_{z_2}(\vec b))}
\end{eqnarray}
and reformulate the integral as
\begin{eqnarray}
  \intop_{-\infty}^{\infty}dz_1\intop_{-\infty}^{\infty}dz_2&=&\intop_{-\infty}^{\infty}dz_1\intop_{z_1}^{\infty}dz_2+\intop_{-\infty}^{\infty}dz_2\intop_{z_2}^{\infty}dz_1\nonumber\\
  &=&2\intop_{-\infty}^{\infty}dz_1\intop_{z_1}^{\infty}dz_2
\end{eqnarray}
where the second equality again follows from the symmetry of the integrand with
respect to the exchange of $z_1$ and $z_2$. The two expressions 
(\ref{eq:falter}) and (\ref{eq:kopeliovich}) are therefore equivalent.

%%%%%%%%%%%%%%%%%%%%%%%%%%%%%%%%%%%%%%%%%%%%%%%%%%%%%%%%%%%%%%%%%%%%%%%%%%%
\newpage
\newpage
\newpage
\newpage

\end{document}